\documentclass[aps,prl,twocolumn,floatfix, showpacs]{revtex4}
\usepackage{graphicx}
\usepackage{dcolumn}
\usepackage{bm}
\usepackage{color}

\usepackage{array}
\usepackage{subfigure}

\newcommand{\be}{\begin{equation}}
\newcommand{\ee}{\end{equation}}
\definecolor{Richard}{rgb}{0,0,1} 
\definecolor{Detlef}{cmyk}{1,0,0,0} 
\definecolor{Guenter}{rgb}{1,0.5,0} 
\definecolor{Jin}{rgb}{1,0,0} 
\definecolor{Herman}{rgb}{0,1,0} 
\definecolor{Roberto}{cmyk}{0,1,0,0} 

\def\gtwid{\mathrel{\raise.3ex\hbox{$>$\kern-.75em\lower1ex\hbox{$\sim$}}}}
\def\ltwid{\mathrel{\raise.3ex\hbox{$<$\kern-.75em\lower1ex\hbox{$\sim$}}}}

\begin{document}

\title{Prandtl-, Rayleigh-, and Rossby-number dependence of heat transport in turbulent rotating Rayleigh-B\'enard convection}
\author{Jin-Qiang Zhong$^1$}
\author{Richard J.A.M. Stevens$^2$}
\author{Herman J.H. Clercx$^3$,$^4$}
\author{Roberto Verzicco$^5$}
\author{Detlef Lohse$^2$}
\author{Guenter Ahlers$^1$}
\affiliation{$^1$Department of Physics and iQCD, University of California, Santa Barbara, CA 93106, USA}
\affiliation{$^2$Department of Science and Technology and J.M. Burgers Center for Fluid Dynamics, University of Twente, P.O Box 217, 7500 AE Enschede, The Netherlands}
\affiliation{$^3$Department of Applied Mathematics, University of Twente, P.O Box 217, 7500 AE Enschede, The Netherlands}
\affiliation{$^4$Department of Physics and J.M. Burgers Centre for Fluid Dynamics, Eindhoven University of Technology, P.O. Box 513, 5600 MB Eindhoven, The Netherlands}
\affiliation{$^5$Dept. of Mech. Eng., Universita' di Roma "Tor Vergata",
Via del Politecnico 1, 00133, Roma.}
\date{\today}

\begin{abstract}
Experimental and numerical data for the heat transfer as a function
of the Rayleigh-, Prandtl-, and Rossby numbers in turbulent
rotating Rayleigh-B\'enard convection are presented.
For relatively small $Ra \approx 10^8$ and large
$Pr$ modest rotation can enhance the heat transfer by
up to 30\%.
At larger $Ra$ there is less heat-transfer enhancement,  and at small $Pr \ltwid 0.7$
there is no heat-transfer enhancement at all. We suggest that the small-$Pr$ behavior is  due to the breakdown of the heat-transfer-enhancing Ekman pumping
because of larger thermal diffusion.
\end{abstract}


\maketitle

Turbulent convection of a fluid contained between two parallel plates and heated from below, known as Rayleigh-B\'enard convection (RBC), continues to be a topic of intense research \cite{Si81,Ka01,AGL09}. A particularly interesting variation of RBC is the case where the sample is rotated about a vertical axis at an angular speed $\Omega$. That system is relevant to numerous astro- and geo-physical phenomena, including convection in the arctic ocean \cite{MS99}, in Earth's outer core \cite{GCHR99}, in the interior of gaseous giant planets \cite{Bu94}, and in the outer layer of the Sun \cite{Mi00}. Thus the problem is of interest in a wide range of sciences, including geology, oceanography, climatology, and astrophysics.

It is widely understood \cite{Ch61} that rotation tends to suppress convective flow, and  with it convective heat transport, when the rate of rotation is sufficiently large.
However, at modest rotation rates, experiments \cite{Ro69,LE97,ZES93}
and numerical simulations \cite{JLMW96,KCG06,OSV07,KCG08}
have shown that under certain conditions
the heat transport can also be {\it enhanced}, before it rapidly decreases for stronger rotation. This enhancement has been ascribed to
Ekman pumping \cite{har95,har00,har02,KCG06}: Due to the
rotation, rising or falling plumes of hot or cold fluid are stretched into vertical vortices that suck fluid out of
the thermal boundary layers (BL) adjacent to the bottom and top plates. This process contributes to the vertical heat flux. 
There are, however, some experiments \cite{pfo84} and numerical simulations \cite{OSV07} that show hardly any enhancement of
the heat flux at modest rotation rates.

In the present paper we determine systematically as a function of the Rayleigh number $Ra$, Prandtl number $Pr$, and
Rossby number $Ro$ (all to be defined below) where the heat-flux enhancement occurs.
We present both experimental measurements and results from direct numerical simulation (DNS). They
 cover different but overlapping parameter ranges and thus complement each other. Where they overlap they agree very well. We find that in certain regimes
the heat-flux enhancement can be as large as 30\%; this raises the possibility of relevance in industrial processes. Even more remarkably, we observe a heretofore unanticipated strong dependence of this enhancement on $Pr$
as well as on $Ra$.

For given aspect ratio $\Gamma \equiv D/L $ ($D$ is the cell diameter and $L$ its height) and given geometry (and here we will only consider cylindrical samples with $\Gamma = 1$),
the nature of RBC is determined by the Rayleigh number $Ra = \beta g \Delta L^3/(\kappa\nu)$
and by the Prandtl number $Pr = \nu /\kappa$. Here, $\beta$ is the thermal expansion coefficient, $g$ the gravitational acceleration, $\Delta = T_b-T_t$ the difference between the imposed temperatures $T_b$ and $T_t$ at the bottom and the top of the sample, respectively, and $\nu$ and $\kappa$ the kinematic viscosity and the thermal diffusivity,
respectively. The rotation rate $\Omega$ (given in rad/s) is used in the form of the Rossby number
$
Ro =  \sqrt{\beta g \Delta /L}/(2\Omega)
$.

The convection apparatus was  described in detail as the ``medium sample" in Ref.~\cite{BFNA05}.
Since the previous measurements \cite{FBNA05} it had been completely dis- and re-assembled.
It had copper top and bottom plates, and a new plexiglas side wall of thickness 0.32 cm was installed for the current project. The sample had a diameter $D = 24.8$ cm and a height $L = 24.8$ cm, yielding $\Gamma = 1.00$.  The apparatus was mounted on a rotating table. We used rotation rates up to 0.3 Hz. Thus the Froude number ${Fr} = \Omega^2 (L/2)/g$ did not exceed 0.05, indicating that centrifugal effects were small. Cooling water for the top plate and electrical leads were brought into and out of the rotating frame using Dynamic Sealing Technologies feed-throughs 
mounted on the rotation axis above the apparatus. All measurements were made at constant imposed $\Delta$ and $\Omega$, and fluid properties were evaluated at $T_m = (T_t+T_b)/2$. Data were taken at 60.00$^\circ$C ($Pr = 3.05,~t_v = L^2/\nu = 1.27\times 10^{5}$ sec), 40.00$^\circ$C ($Pr = 4.38,~t_v=9.19\times 10^{4}$ sec), 24.00$^\circ$C ($Pr = 6.26,~t_v = 6.69 \times 10^{4}$ sec) and 23.00$^\circ$C ($Pr = 6.41,~t_v = 6.56 \times 10^{4}$ sec). In a typical run the system was allowed to equilibrate for three or four hours, and temperatures and heat currents were then averaged over an additional three or four hours and used to calculate $Ra$ and the Nusselt number $Nu = Q L / ( \lambda \Delta)$ ($Q$ is the heat-current density).

\begin{figure}
\includegraphics[width=3.25in]{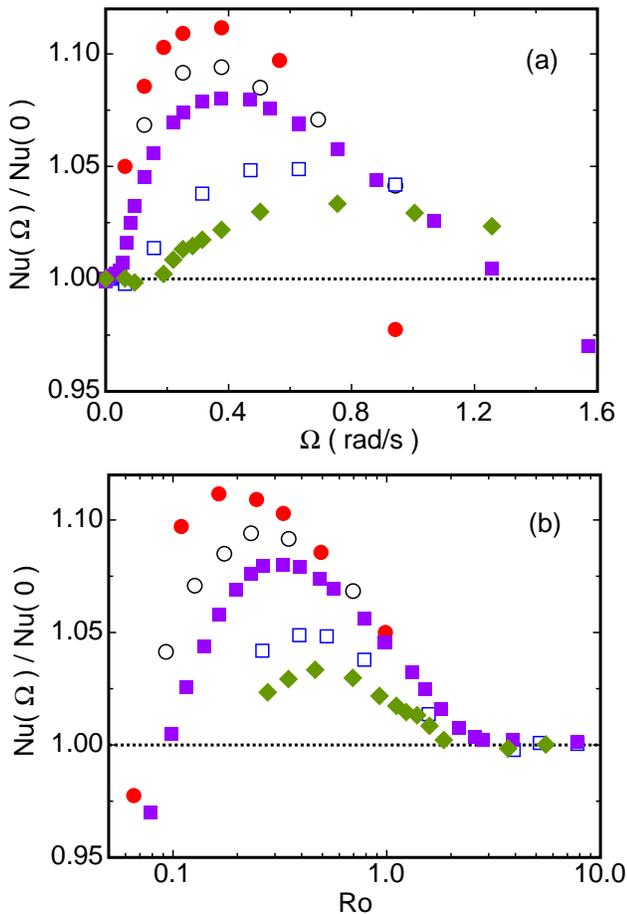}
\caption{The ratio of the Nusselt number $Nu(\Omega)$ in the presence of rotation to $Nu(\Omega=0)$ for $Pr=4.38$ ($T_m = 40.00^\circ$C). (a): Results as a function of the rotation rate in rad/sec. (b): The same results as a function of the Rossby number $Ro$ on a logarithmic scale. Red solid circles: $Ra = 5.6\times 10^8$ ($\Delta = 1.00$ K). Black open circles: $Ra = 1.2\times 10^9$  ($\Delta = 2.00$ K). Purple solid squares: $Ra = 2.2\times 10^9$  ($\Delta = 4.00$ K). Blue open squares: $Ra = 8.9\times 10^9$  ($\Delta = 16.00$ K). Green solid diamonds: $Ra = 1.8\times 10^{10}$  ($\Delta = 32.00$ K)}
\label{fig:Nu_1}
\end{figure}

\begin{figure}
\includegraphics[width=3.25in]{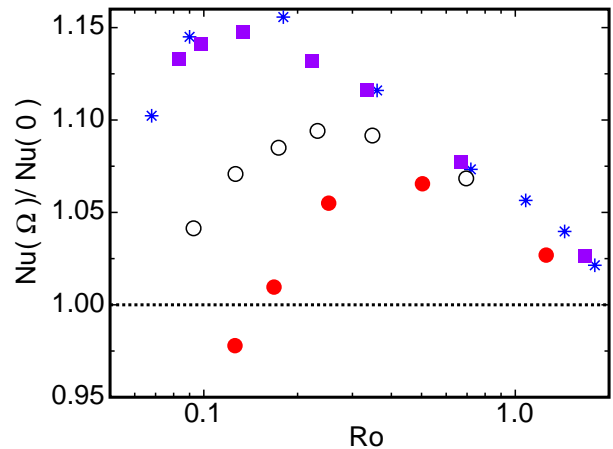}
\caption{The ratio $Nu(\Omega)/Nu(\Omega=0)$ for $Ra = 1.2\times 10^9$ as function of $Ro$ on a logarithmic scale. Red solid circles: $Pr = 3.05$ ($T_m = 60.00^\circ$C). Black open circles: $Pr = 4.38$  ($T_m = 40.00^\circ$C). Purple solid squares: $Pr = 6.41$  ($T_m = 23.00^\circ$C). Blue stars: Numerical simulation for $Ra = 1.0\times 10^9$ and $Pr = 6.4$ from Ref.~\cite{KCG08}.}
\label{fig:Nu_2}
\end{figure}

Measurements of the Nusselt number without rotation, $Nu(\Omega=0)$, over the range $5\times 10^8 \alt Ra \alt 10^{10}$ ($1 \alt \Delta \alt 20$K) agreed within estimated systematic errors of about 1\%
 with previous results \cite{FBNA05} obtained in the same apparatus. The ratio $Nu(\Omega)/Nu(\Omega=0)$ is shown in Fig.~\ref{fig:Nu_1}a as a function of the rotation rate $\Omega$. Those results are for $T_m = 40.00^\circ$C, where $Pr = 4.38$. The enhancement of $Nu$ due to modest rotation is clearly seen at all $Ra$. It is larger at the smaller $Ra$.

In Fig.~\ref{fig:Nu_1}b we show the same data as a function of the Rossby number $Ro$. At large $Ro$ (small $\Omega$) the data must approach unity, and indeed they do. As $Ro$ decreases, $Nu$ is first enhanced, but then reaches a maximum and decreases as expected. The maximum of $Nu(\Omega)/Nu(\Omega=0)$ occurs at larger $Ro$ for larger $Ra$, and the value of this ratio at the maximum diminishes with increasing $Ra$.

In Fig.~\ref{fig:Nu_2} the results at constant $Ra$ are shown as a function of $Ro$ for several values of $Pr$. Also shown are the DNS results from Ref.~\cite{KCG08}; these data agree rather well with the experimental data at nearly the same $Pr$ and $Ra$. One sees that the enhancement of $Nu$ at large $Ro$ is nearly independent of $Pr$; but as $Ro$ decreases below unity a strong $Pr$ dependence manifests itself. As $Ro$ decreases, the depression of $Nu$ sets in earlier for smaller $Pr$, and the maximal relative heat transfer enhancement is smaller.

\begin{figure}
\includegraphics[width=3.25in]{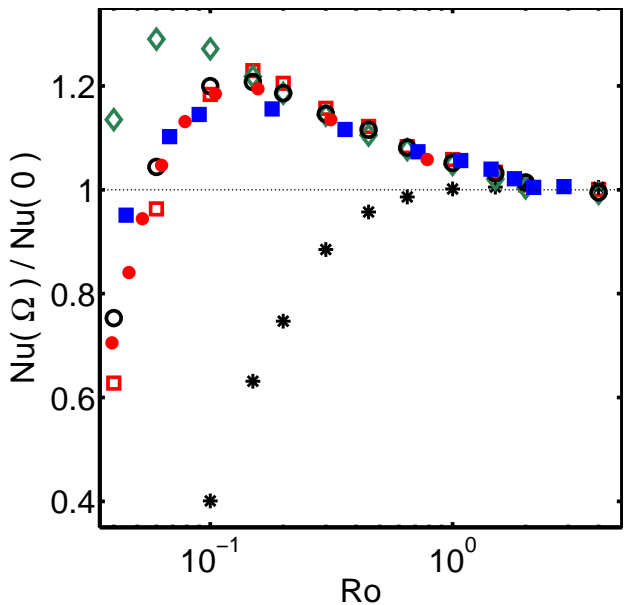}
\caption{The ratio $Nu(\Omega)/Nu(\Omega=0)$ as function of $Ro$ on a logarithmic scale. Red solid circles: $Ra = 2.73\times 10^8$ and $Pr = 6.26$ (experiment). Black open circles: $Ra = 2.73\times 10^8$ and $Pr = 6.26$ (DNS). Blue solid squares: $Ra = 1\times 10^9$ and $Pr = 6.4$ (DNS) \cite{KCG08}. Red open squares: $Ra = 1\times 10^8$ and $Pr = 6.4$ (DNS). Green open diamonds: $Ra = 1\times 10^8$ and $Pr = 20$ (DNS). Black stars: $Ra = 1\times 10^8$ and $Pr = 0.7$ (DNS). 
}
\label{fig:Nu_3}
\end{figure}

In the DNS we solved the three-dimensional Navier-Stokes equations within the Boussinesq approximation,
\begin{eqnarray}
 \frac{D\textbf{u}}{Dt} &=& - \nabla P + \left( \frac{Pr}{Ra} \right)^{1/2} \nabla^2 \textbf{u} + \theta \textbf{$\widehat{z}$}- \frac{1}{Ro} \widehat{z} \times \textbf{u}, \\
 \frac{D\theta}{Dt} &=& \frac{1}{(PrRa)^{1/2}}\nabla^2 \theta , 
\end{eqnarray}
with
 $\nabla \cdot \textbf{u} = 0$.
 Here \textbf{$\widehat{z}$} is the unit vector pointing in the opposite direction to gravity, $\textbf{u}$ the velocity vector, and $\theta$ the non-dimensional temperature, $0\leq \theta \leq 1$.
 Finally, $P$ is the reduced pressure (separated from its hydrostatic contribution, but containing the centripetal contributions): $P=p - r^2/(8Ro^2)$, with $r$ the distance to the rotation axis. The equations have been made non-dimensional by using, next to $L$ and $\Delta$,
 the free-fall 
 velocity $U=\sqrt{\beta g \Delta L}$.
 The simulations were performed on a grid of $129 \times 257 \times 257$ nodes, respectively, in the radial, azimuthal and vertical directions, allowing for a sufficient resolution of the small scales both inside the bulk of turbulence and in the BLs
(where the grid-point density has been enhanced) for the parameters employed here
\cite{OSV07,KCG08}. $Nu$ is calculated as in Refs.~\cite{OSV07,KCG08} and its statistical convergence has been controlled.

The numerical results for $Nu$ as function of $Ro$ for several $Ra$ and $Pr$ are shown in Fig.~\ref{fig:Nu_3}. For $Ra = 2.73\times 10^8$ and $Pr = 6.26$  experimental data (not previously shown in Figs.~\ref{fig:Nu_1} and \ref{fig:Nu_2}) are plotted as well,
showing near-perfect agreement with the numerical results. This gives us confidence that the partial neglect of centrifugal effects in the simulations [namely, neglecting the density dependence
of the centripetal forces, which in the Boussinesq equations show up as $-2 \hbox{Fr} ~ r\theta \widehat{r}$ \cite{HH69}
(with the radial 
unit vector $\widehat{r}$ )]
is justified, as already found in
\cite{KCG06,KCG08}, because $\hbox{Fr}\ll 1$.
 Neither do  the experimentally unavoidable
finite conductivity of the top and bottom plates \cite{Ve04,BFNA05} and the side-wall conductivity \cite{Ah00,RCCHS01} seem to matter in this regime of parameter space,
as already explained in Ref.~\cite{AGL09} for the non-rotating case.

\begin{figure}[!hbt]
\includegraphics[width=3.25in]{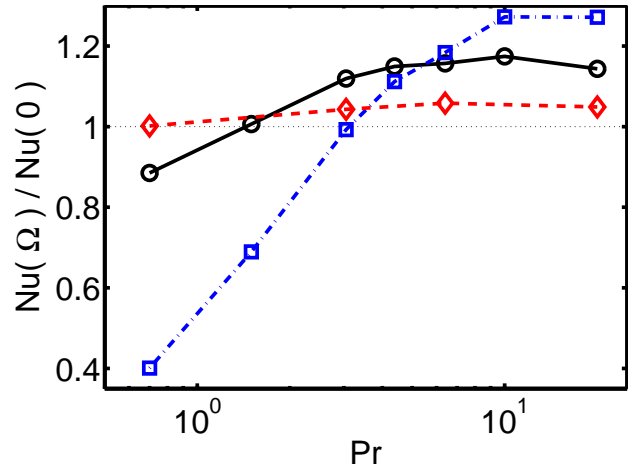}
\caption{Numerical result for the ratio $Nu(\Omega)/Nu(0)$ as function of $Pr$ for $Ra =  10^8$ and $Ro=1.0$ (red open diamonds), $Ro=0.3$ (black open circles), and $Ro=0.1$ (blue open squares).}
\label{fig:Nu_4}
\end{figure}

As was seen for the experimental data in Figs.~\ref{fig:Nu_1} and \ref{fig:Nu_2}, the numerical results in Fig.~\ref{fig:Nu_3} also reveal a drastic dependence of the Nusselt-number enhancement on $Pr$.
For large $Pr\gtwid 6$ the enhancement can be as large as 30\% for $Ra \approx 10^8$ and $Ro \approx 0.1$.
 However, there is no enhancement at all for small $Pr \ltwid 0.7$. This trend
is further elucidated in Fig.~\ref{fig:Nu_4}, where we show
the Nusselt-number enhancement as function of $Pr$ for three different $Ro$ and $Ra = 10^8$.

\begin{figure}
\centering
\subfigure{\includegraphics[width=0.21\textwidth]{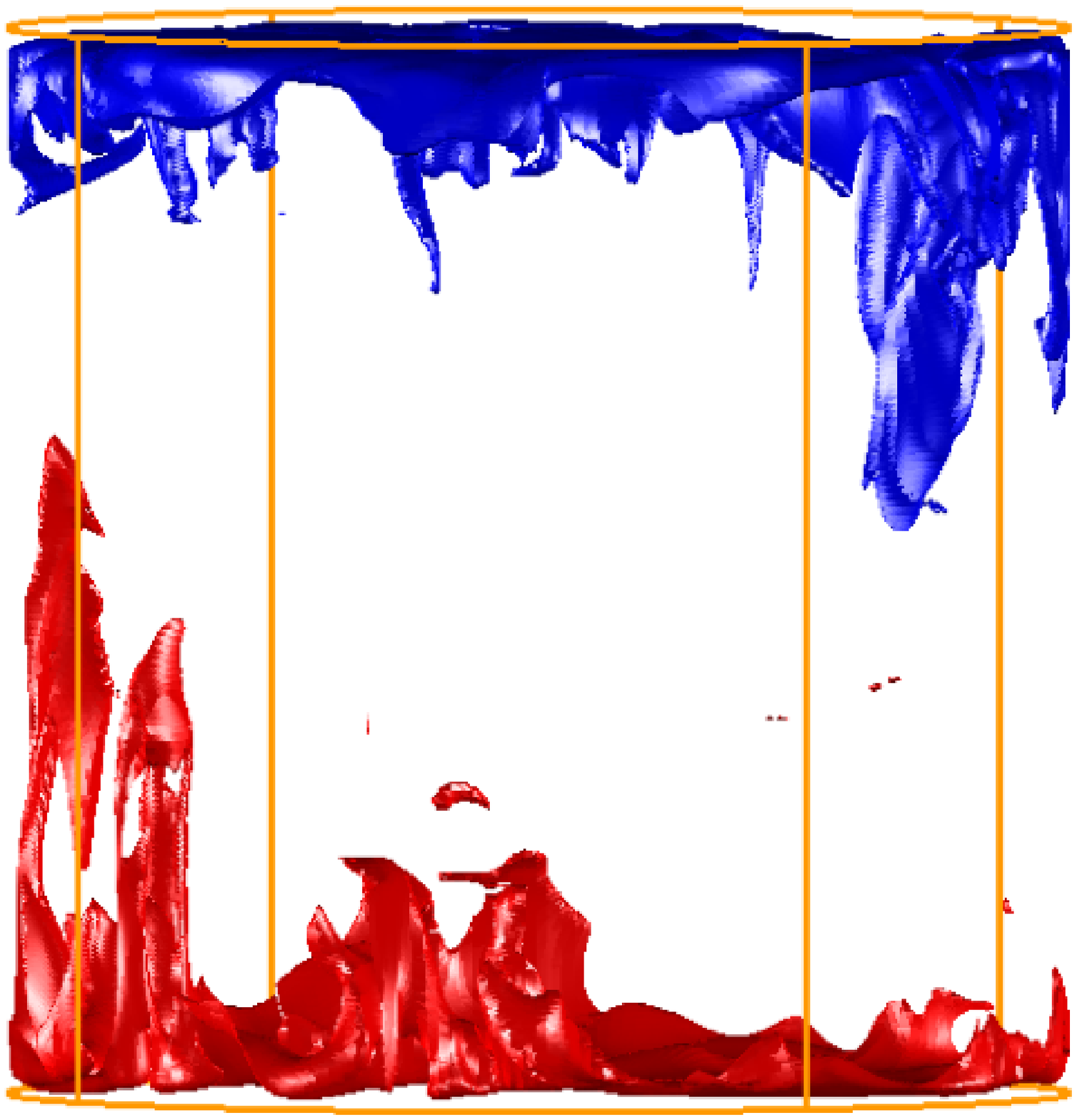}}
\hspace{3mm}
\subfigure{\includegraphics[width=0.21\textwidth]{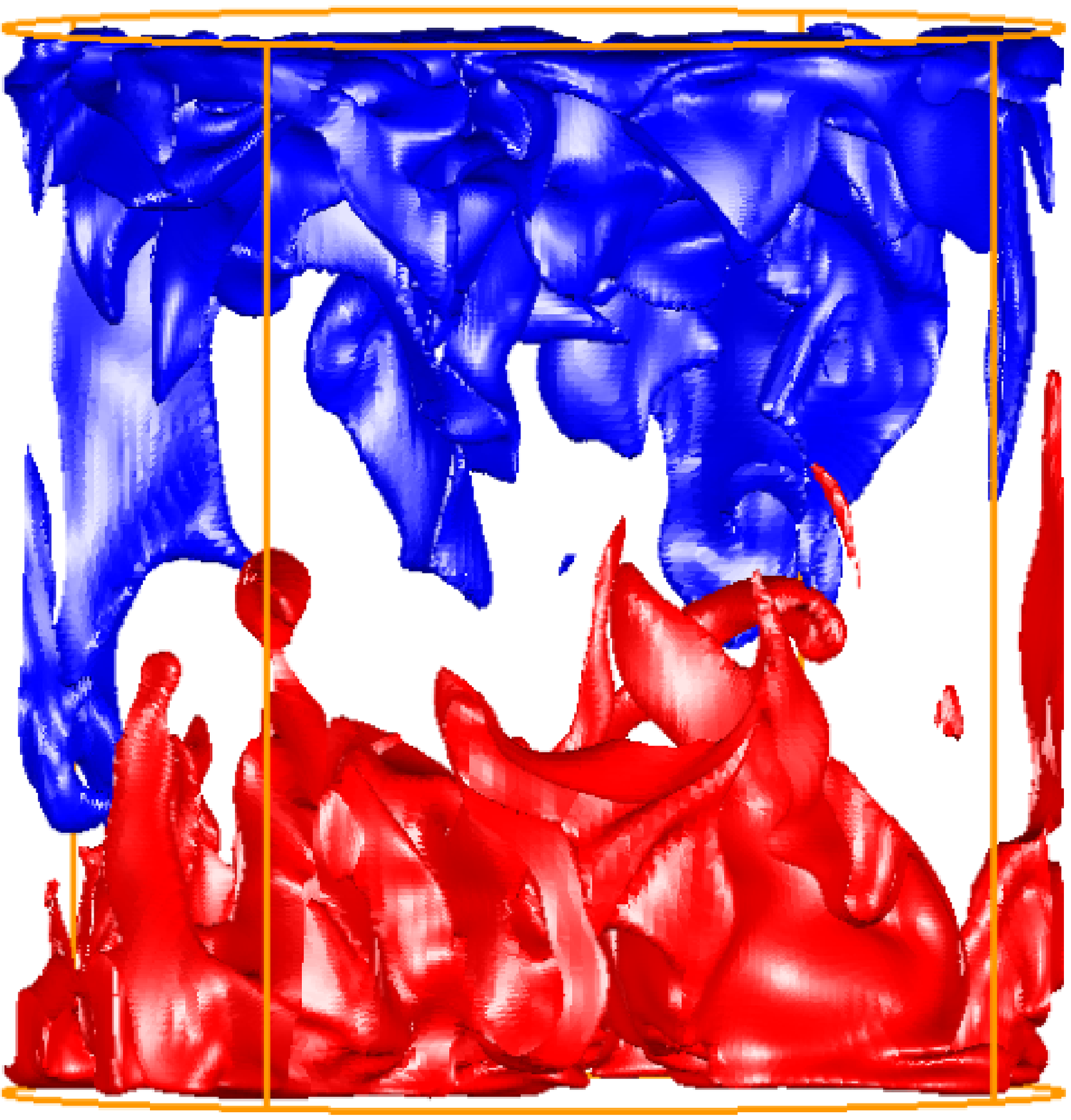}}
\vspace{3mm}
\subfigure{\includegraphics[width=0.21\textwidth]{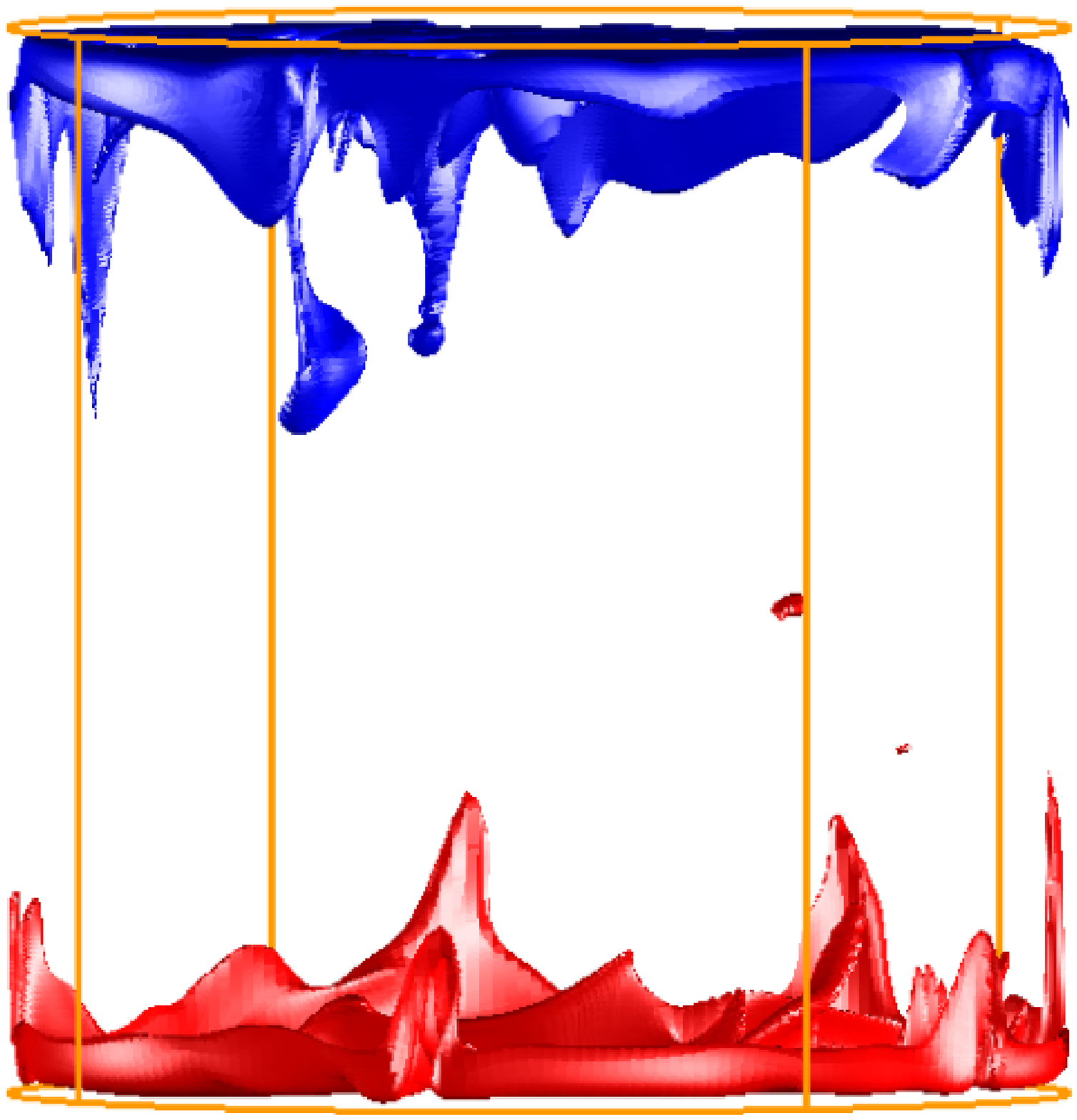}}
\hspace{3mm}
\subfigure{\includegraphics[width=0.21\textwidth]{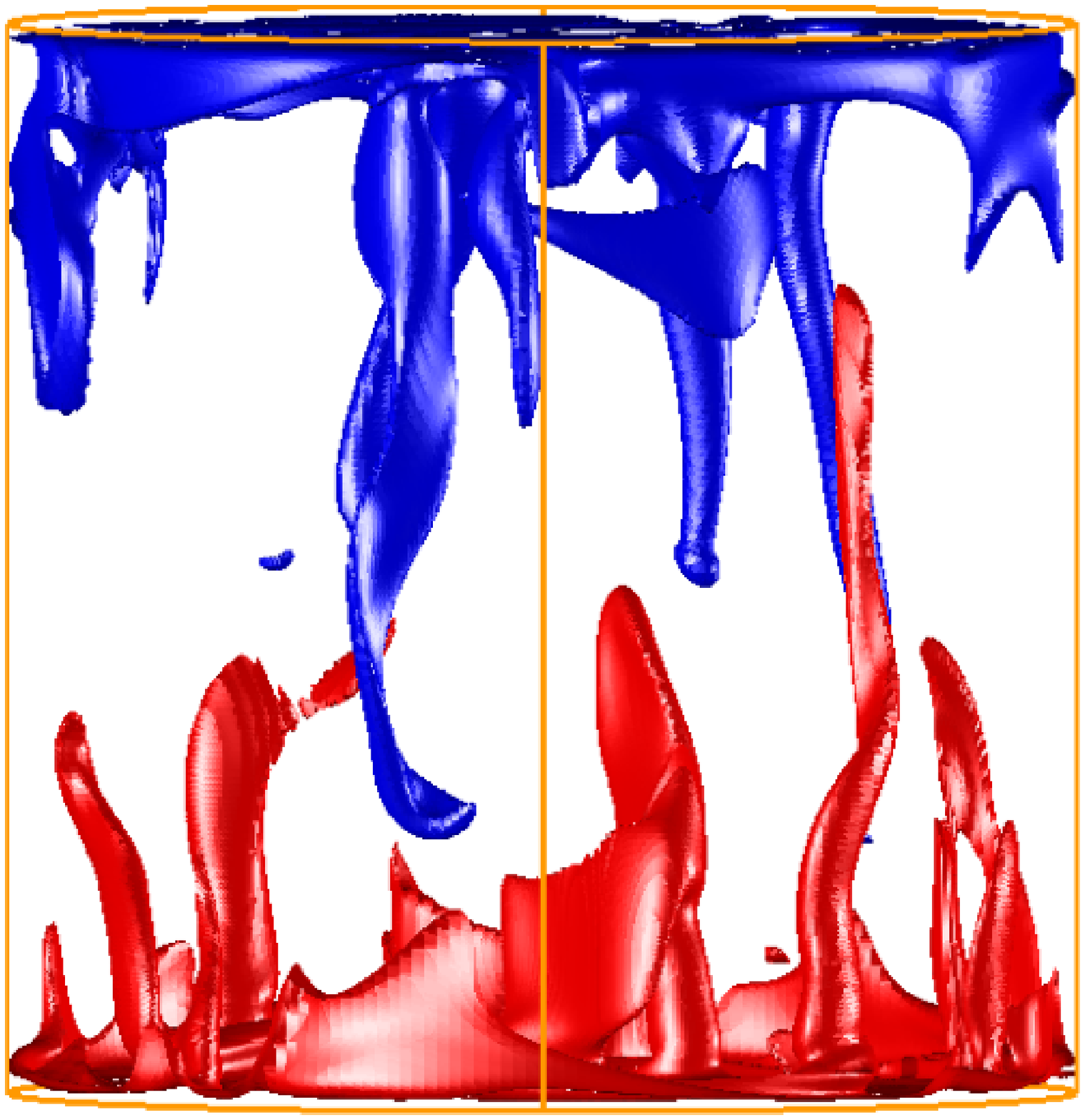}}
\caption{3D visualization of the temperature isosurfaces at
$0.65\Delta$ (red) and $0.35\Delta$ (blue), respectively,
for $Pr = 0.7$ (upper plot) and $Pr = 6.4$ (lower plot) for
 $Ra =  10^8$ and $Ro=\infty$(left) $Ro=0.30$ (right).}
\label{fig:3D-plots}
\end{figure}

\begin{figure}[!hbt]
\includegraphics[width=3.25in]{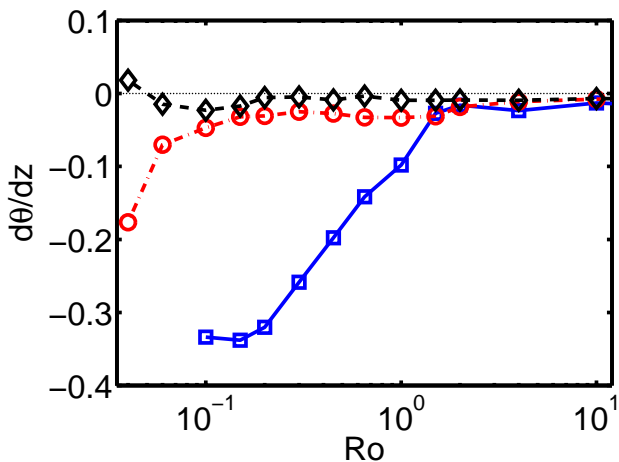}
\caption{The horizontally averaged vertical temperature gradient $d\theta/dz$ at the sample mid-plane as a function of $Ro$ for $Ra = 1\times 10^8$. Blue open squares: $Pr = 0.7$. Red open circles: $Pr = 6.4$. Black open diamonds: $Pr = 20$. }
\label{fig:dTdz}
\end{figure}

Why does the heat-transfer enhancement through Ekman pumping
at modest rotation rates break down as $Ro$ decreases  
below a $Pr$-dependent typical value, as seen in Figs.\  \ref{fig:Nu_1} to \ref{fig:Nu_3}?
To obtain a hint,
we visualized (see Fig.~\ref{fig:3D-plots}) the three-dimensional temperature iso-surfaces
for $Pr = 0.7$ and for $Pr = 6.4$ at both
 $Ro=0.30$ and $Ro=\infty$,  at  $Ra = 10^8$.
While for the larger $Pr = 6.4$ case the temperature
iso-surfaces organize to reveal long vertical vortices as suggested by the
Ekman-pumping picture, these structures are much shorter and
broadened for the low $Pr = 0.7$ case, due to the larger thermal diffusion
which makes the Ekman pumping inefficient.
This would imply enhanced horizontal heat transport which should also lead to a steeper gradient of the mean temperature in the bulk.
Indeed, 
in the DNS, we find  that when 
$Ro$ becomes small enough, 
the bulk of the fluid displays an increasingly 
destabilizing mean temperature gradient (see Fig.~\ref{fig:dTdz}), which of course must be 
accompanied by a reduction of the mean temperature drop over the thermal BLs and thus 
a Nusselt-number reduction. The first manifestation of the enhancement of the mean destabilizing vertical temperature gradient  agrees with the onset of relative heat-transfer reduction in figure \ref{fig:Nu_3}, and thus supports this 
 explanation.

Along the same line
of arguments one may also expect that the $Ra$ dependence of the reduction of $Nu$ at small $Ro$ seen in Fig.~\ref{fig:Nu_1} is attributable to relatively less efficient Ekman pumping at
higher $Ra$: The enhanced turbulence
may lead to a larger eddy thermal diffusivity, promoting a homogeneous mean temperature in the bulk. Again this would make Ekman pumping relatively less
efficient and reduce the peak in the relative Nusselt number.

Another interesting aspect of our data is that within our experimental or numerical resolution there is no heat-flux enhancement for $Ro \gtwid 2$ for any
$Ra$ or $Pr$. As already noticeable from the data of
Ref.\ \cite{KCG08}, the heat-flux enhancement first becomes resolved as $Ro$ decreases below about two. Future work has to reveal whether this onset coincides
with some reorganization of the flow. Similarly, 
it remains to be
analyzed how the dramatic dependence of $Nu$ on $Ra$ and $Pr$ for modest $Ro$ is reflected in the flow fields.

{\it Acknowledgements:}
The experimental work was supported by the U.S. National Science Foundation through Grant DMR07-02111. RJAMS wishes to thank the Foundation for Fundamental Research on Matter (FOM) for financial support. This work was sponsored by the National Computing Facilities Foundation (NCF) for the use of supercomputer facilities, with financial support from the Netherlands Organization for Scientific Research (NWO).


\end{document}